\documentclass[12pt,twoside,journal,onecolumn,a4paper]{IEEEtran}

\usepackage{amssymb, amsmath}

\newtheorem{theorem}{Theorem}
\newtheorem{definition}[theorem]{Definition}
\newtheorem{algorithm}[theorem]{Algorithm}

\newcommand{\beqn}{\begin{equation}}
\newcommand{\eeqn}{\end{equation}}
\newcommand{\beq}{\begin{equation*}}
\newcommand{\eeq}{\end{equation*}}
\newcommand{\Z}{\mathbb Z}
\newcommand{\C}{\mathbb C}
\renewcommand{\H}{{\scriptscriptstyle H}}
\renewcommand{\th}{{\rm th}}
\newcommand{\RM}{{\rm RM}}
\renewcommand{\Re}{{\rm Re}}

\newcommand{\bix}{{\boldsymbol{x}}}
\newcommand{\biy}{{\boldsymbol{y}}}
\newcommand{\biw}{{\boldsymbol{w}}}
\newcommand{\bic}{{\boldsymbol{c}}}
\newcommand{\biu}{{\boldsymbol{u}}}
\newcommand{\biz}{{\boldsymbol{z}}}
\newcommand{\bin}{{\boldsymbol{n}}}

\newcommand{\bG}{{\mathbf{G}}}
\newcommand{\bnull}{{\boldsymbol{0}}}
\newcommand{\bone}{{\boldsymbol{1}}}

\begin{document}

\date{}
\title{Simple Maximum-Likelihood Decoding of Generalized First-order Reed--Muller Codes}%

\author{Kai-Uwe Schmidt and Adolf Finger
\thanks{This paper is accepted for publication in IEEE Communications Letters.}%
\thanks{K.-U. Schmidt and A. Finger are with Communications Laboratory, Technische Universit\"at Dresden, 01062, Dresden, Germany (e-mail: schmidtk@ifn.et.tu-dresden.de)}%
}
\maketitle

\begin{abstract}
An efficient decoder for the generalized first-order Reed--Muller code $\RM_q(1,m)$ is essential for the decoding of various block-coding schemes for orthogonal frequency-division multiplexing with reduced peak-to-mean power ratio. We present an efficient and simple maximum-likelihood decoding algorithm for $\RM_q(1,m)$. It is shown that this algorithm has lower complexity than other previously known maximum-likelihood decoders for $\RM_q(1,m)$.
\end{abstract}

\section{Introduction} 
\PARstart{I}{n order} to ensure tight control of the peak-to-mean power ratio (PMPR) in orthogonal frequency-division multiplexing (OFDM) systems, various block-coding schemes have been proposed \cite{Davis1999}, \cite{Paterson2000a}. These codes are obtained from unions of, say $M$, cosets of a $q$-ary generalization of the first-order Reed--Muller code $\RM_q(1,m)$ (see Definition \ref{def:RM}). Such codes have the potential to perform error correction and ensure a substantially reduced PMPR compared to uncoded transmission. 
\par
Most of the existing decoding algorithms for the above mentioned codes are based on the supercode decoding method, described in \cite{Conway1986} for the binary case. Such a decoding scheme involves $M$ decodings of $\RM_q(1,m)$. Hence an efficient decoder for $\RM_q(1,m)$ is required. Moreover, if a maximum-likelihood (ML) decoder is used to decode $\RM_q(1,m)$, then the supercode decoder performs ML decoding too \cite{Conway1986}, \cite{Davis1999}.
\par
There exist various decoding techniques for the code $\RM_2(1,m)$. The fast Hadamard transform (FHT) can be used to obtain an ML decoder \cite{MacWilliams1977}. Suboptimal techniques include majority-logic decoding \cite{MacWilliams1977} and decoding based on the interpretation of the Reed--Muller codes as general concatenated codes \cite{Schnabl1995}, \cite{Dumer2004}. In \cite{Lucas1998a} the latter approach was extended to a list-decoding scheme, and it was shown that ML decoding of $\RM_2(1,m)$ is possible if the list length is equal to 2.
\par
In \cite{Davis1999} an iterative decoder for $\RM_{2^h}(1,m)$ was proposed that relegates the decoding of $\RM_{2^h}(1,m)$ to $h$ decodings of $\RM_2(1,m)$, for which efficient methods exist. A quite similar approach has been reported in \cite{Grant1998a}. Another suboptimal decoder was proposed in \cite{Paterson2000b}, where the majority-logic decoding method was extended to the nonbinary case. Implicitly, a decoder for $\RM_q(1,m)$ was obtained in \cite{Ochiai2000}, by applying the method of ordered statistics to nonbinary codes. In \cite{Jones1999} an ML decoder was given for $\RM_4(1,m)$, by treating $\RM_4(1,m)$ itself as a union of $2^m$ cosets of $\RM_2(1,m+1)$. A $q$-ary equivalent of the binary FHT has been reported in \cite{Grant1998a} and \cite{Grant1998}. Consequently an ML decoder for $\RM_q(1,m)$ for arbitrary $q$ was obtained. In this letter we present a new ML decoder for $\RM_q(1,m)$, which appears to have lower complexity than the above mentioned ML decoding schemes.

\section{Definitions and Preliminaries}
\label{sec:definitions}
\begin{definition}
\label{def:RM}
The generalized first-order Reed--Muller code $\RM_q(1,m)$ ($m\ge1$) consists of the $\Z_q$-valued codewords of length $2^m$ $\{\biu\cdot\bG_m\pmod q\,|\,\biu\in\Z_q^{m+1}\}$, where we define a generator matrix $\bG_m$ recursively by
\beq
\bG_m=
\begin{pmatrix}
\bG_{m-1} & \bG_{m-1}\\
\bnull_{2^{m-1}}    & \bone_{2^{m-1}}
\end{pmatrix}\quad \mbox{with }\quad 
\bG_1=
\begin{pmatrix}
1 & 0\\
0 & 1 
\end{pmatrix}.
\eeq
Here $\bnull_n$ and $\bone_n$ are row vectors of length $n$ containing only zeros and ones, respectively.
\end{definition}
\par
The above definition implies the following recursive construction of the codewords of $\RM_q(1,m)$
\beq
\biu\cdot\bG_m\,(\!\bmod q)=(\biu'\cdot\bG_{m-1}|\biu'\cdot\bG_{m-1}+u_{m}\bone_{2^{m-1}})\,(\!\bmod q).
\eeq
Here $(\cdot|\cdot)$ means concatenation and $\biu=(u_0\,u_1\cdots u_{m})$ and $\biu'=(u_0\,u_1\cdots u_{m-1})$.
This property will be used in the next section to derive an efficient ML decoding scheme for $\RM_q(1,m)$.
\par
After encoding, the $\Z_q$-valued codewords are mapped onto a $q$-ary phase-shift-keying (PSK) constellation. This means that for each $\Z_q$-valued codeword there exists a corresponding polyphase codeword, which is given by
\beq
\xi^{\biu\cdot\bG_m}=\xi^{\bic}=(\xi^{c_0}\,\xi^{c_1}\cdots\xi^{c_{{2^m}-1}}).
\eeq
Here $\xi=\exp(j2\pi/q)$ is a primitive $q\th$ root of unity, where $j^2=-1$. Since there is a one-to-one correspondence between $\Z_q$-valued and polyphase codewords, in the remainder of this letter, we shall drop the distinction between them. The context should make clear to which one we refer.

\section{Decoding of $\RM_q(1,m)$}

We assume that the codeword $\xi^{\biu\cdot\bG_m}$ is transmitted over an additive white Gaussian noise (AWGN) channel. Then we receive
\beq
\biy=\xi^{\biu\cdot\bG_m}+\bin,
\eeq
where $\bin\in\C^{2^m}$ is the complex white Gaussian noise vector. An ML decoder now finds the codeword $\hat\bix$ that is closest to $\biy$ in the Euclidean sense. We shall call $\hat\bix$ the ML codeword of $\biy$. That is to say
\beq
\hat\bix=\arg\min_{\bix\in\RM_q(1,m)}||\biy-\bix||^2,
\eeq
where
\beq
||\biy-\bix||^2=\sum_{k=0}^{2^m-1}|y_k-x_k|^2.
\eeq
It is straightforward to verify 
\beq
||\biy-\bix||^2=||\biy||^2+||\bix||^2-2\,\Re\{\biy\cdot\bix^\H\},
\eeq
where $()^\H$ denotes Hermitian (conjugate transpose), such that $\biy\cdot\bix^\H$ is the dot product of $\biy$ and $\bix$.   Notice that $||\biy||^2$ and $||\bix||^2$  are independent of the decoding result, since $||\bix||^2=n$ for each $\bix\in\RM_q(1,m)$. Hence finding the codeword that is closest to $\biy$ is equivalent to finding the codeword for which the real part of the dot product with the received vector is maximized. To be precise
\beq
\hat\bix=\arg\max_{\bix\in\RM_q(1,m)}\Re\{\biy\cdot\bix^\H\}.
\eeq
\par
We now state our decoding algorithm. Then we show that this algorithm always outputs the ML codeword of $\biy$.
\begin{algorithm}
\label{alg:Dec}
Soft-decision maximum-likelihood decoder for $\RM_q(1,m)$.
\begin{enumerate}
\item Input a vector $\biy=(y_0\,y_1\cdots y_{2^m-1})\in\C^{2^m}$.
\item If $m=1$, output the hard decisions $\hat\bix$, the estimated information symbols $\hat\biu$, and $\Re\{\biy\cdot\hat\bix^\H\}$. Stop.
\item For $i=0,1,\cdots,q-1$ calculate
\beq
\biz(i)=(z_0(i)\,z_1(i)\cdots z_{2^{m-1}-1}(i))
\eeq
with 
\beq
z_k(i)=y_k+y_{k+2^{m-1}}\xi^{-i},
\eeq
and use the algorithm to decode $\biz(i)$ to $\hat\biz(i)$, to get the estimated information symbols $\hat\biu'(i)$, and to calculate $p(i)=\Re\{\biz(i)\cdot\hat\biz^\H(i)\}$.
\item Determine 
\beq
\hat \imath=\arg\max\limits_{i\in\Z_q}p(i).
\eeq 
Output the decoded codeword $\hat\bix=(\hat\biz(\hat \imath)|\hat\biz(\hat \imath)\xi^{\hat \imath})$, the estimated information symbols $\hat\biu=(\hat\biu'(\hat\imath)|\hat\imath)$, and $p(\hat \imath)$.
\end{enumerate}
\end{algorithm}
\par
\begin{theorem}
Algorithm \ref{alg:Dec} is a soft-decision maximum-likelihood decoder for $\RM_q(1,m)$.
\end{theorem}
\par
\begin{proof}
We will show that, if Algorithm \ref{alg:Dec} is an ML decoder for $\RM_q(1,m-1)$, then it is also an ML decoder for $\RM_q(1,m)$. The  case $m=1$ serves as the induction anchor, since it is obvious that Algorithm \ref{alg:Dec} is an ML decoder for $\RM_q(1,1)$. Moreover it is clear that, if $m=1$, Algorithm \ref{alg:Dec} outputs $\Re\{\biy\cdot\hat\bix^\H\}$ correctly.
\par
Now let $m>1$. We have to show that the algorithm always outputs the ML codeword $\hat\bix$ and $\Re\{\biy\cdot \hat\bix^\H\}$, which is potentially required for the higher stage of the decoder. From the discussion in Section \ref{sec:definitions} we know that a polyphase codeword $\bix\in\RM_q(1,m)$ may be expressed as $\bix=(\biw|\biw\,\xi^{i})$, where $\biw\in\RM_q(1,m-1)$ and $i\in\Z_q$. 
By computing
\beqn
(\hat\biw,\hat\imath)=\arg\max_{\biw\in\RM_q(1,m-1),i\in\Z_q}\Re\{\biy\cdot( \biw|\biw\,\xi^{i})^\H\},
\label{eqn:argmax}
\eeqn
we can obviously find $\hat\bix=(\hat\biw|\hat\biw\,\xi^{\hat\imath})$. Let us inspect the real part of the dot product in (\ref{eqn:argmax})
\beqn
\Re\{\biy\cdot( \biw|\biw\,\xi^{i})^\H\}\\
\quad=\sum_{k=0}^{2^{m-1}-1}\Re\{(y_k+y_{k+2^{m-1}}\xi^{-i}) w_k^*\}=\Re\{\biz(i)\cdot\biw^\H\},
\label{eqn:split-dot-prod}
\eeqn
where $()^*$ denotes complex conjugation and $\biz(i)$ is as calculated in Step 3 of Algorithm \ref{alg:Dec}. By hypothesis Algorithm \ref{alg:Dec} is an ML decoder for $\RM_q(1,m-1)$, which is now used to get $\hat\biz(i)$ according to
\beqn
\hat\biz(i)=\arg\max_{\biw\in\RM_q(1,m-1)}\Re\{\biz(i)\cdot\biw^\H\}.
\label{eqn:max-dot-z}
\eeqn
Moreover the algorithm provides $p(i)=\Re\{\biz(i)\cdot(\hat\biz(i))^\H\}$. In order to find $\hat\biw$, and thus also $\hat\bix$, it remains to compute $\hat \imath=\arg\max_{i\in\Z_q}p(i)$, which is done in Step 4 of Algorithm \ref{alg:Dec}. Then we have $\hat\biw=\hat\biz(\hat\imath)$ and $\hat\imath$, from which $\hat\bix$ can be obtained. We also have, by (\ref{eqn:split-dot-prod}) and (\ref{eqn:max-dot-z}),  $p(\hat\imath)=\Re\{\hat\biz(\hat\imath)\cdot(\hat\biz(\hat\imath))^\H\}=\Re\{\biy\cdot\hat\bix^\H\}$.
\end{proof}
\par
We close this section with a brief discussion. 
The code $\RM_q(1,m)$ may be interpreted as a union of $q$ cosets of the code $\{(\biw|\biw)\,|\,\biw\in\RM_q(1,m-1)\}$ with coset representatives $\{i\cdot(\bnull_{2^{m-1}}|\bone_{2^{m-1}})\,|\,i\in\Z_q\}$. Then Algorithm \ref{alg:Dec} basically performs the steps of the supercode decoding principle, as stated in \cite{Conway1986} for binary codes. We also mention the relationship of Algorithm \ref{alg:Dec} and the list-decoding scheme, proposed in \cite{Lucas1998a}. In this reference it was shown that an ML decoder for $\RM_2(r,m)$ can be obtained when the list length is equal to 2. The basic idea behind Algorithm \ref{alg:Dec} can therefore also be interpreted as a generalization of the list-decoding method for nonbinary codes, where the list length is set to $q$.

\section{Complexity Analysis and Comparisons}

We next analyze the complexity of Algorithm \ref{alg:Dec} and make comparisons with existing approaches. In practice it is likely that $q$ is a power of 2. So we restrict our analysis to the case $q=2^h$. We consider the number of complex additions $N^+(2^h,m)$ and the number of complex multiplications $N^\times(2^h,m)$ as complexity measurements. Notice that we regard one real addition as half a complex addition. Multiplications with $1,j,-1,-j$ are not counted, since they are trivial and can be implemented by sign bit changes and swapping of the real and imaginary part.
\par
We start with $m=1$. Then, in Step 2, we need to determine the hard decisions and the dot product of two vectors of length 2. The hard decision for the $k\th$ symbol is $\hat x_k=\xi^{\hat u_k}$, where
\beqn
\hat u_k=\arg\max_{i\in\Z_{2^h}}\Re\{y_k\cdot\xi^{-i}\}.
\label{eqn:det_hd}
\eeqn
For $h\le 2$ the above calculation is a trivial task and can be accomplished with simple sign logic. For $h>2$ we may implement this maximum calculation as follows. First determine the quadrant in which the received value lies. This leaves $2^{h-2}+1$ candidates, over which the maximum in (\ref{eqn:det_hd}) has to be calculated. For each candidate one complex multiplication is required. Since one of the possible signal points lies on the real and another on the imaginary axis (where the multiplication is trivial), we need for each of the two hard decisions $2^{h-2}-1$ complex multiplications. Hence 
\beq
N^\times(2^h,1)=
\begin{cases}
0,&h\le 2\\
2^{h-1}-2,&h>2.
\end{cases}
\eeq
Having $\Re\{y_k\xi^{-\hat u_k}\}$ for $k=0,1$, one real addition is required to calculate the dot product in Step 2.
Hence
\beq
N^+(2^h,1)=1/2.
\eeq
\par
Now let $m>1$. In Step 3 we require $2^h2^{m-1}$ complex multiplications to calculate $z_k(i)$ for $k=0,\cdots,2^{m-1}-1$ and $i=0,\cdots,2^h-1$. If $h\le 2$, these are trivial multiplications. For $h>2$, 4 out of the $2^h$ complex multiplications for each $k$ are trivial. Moreover, if we have computed $y_k\xi^{-i}$ for $i=0,1,\cdots,2^{h-1}-1$, by sign bit changes, the values $y_k\xi^{-i}$ for $i=2^{h-1},\cdots,2^{h}-1$ can be easily derived. After having the products $y_k\xi^{-i}$, there are $2^h\,2^{m-1}$ additions required in Step 3. These are real additions if $h=1$ and complex additions if $h>1$. Finally we need to perform $2^h$ decodings of $\RM_q(1,m-1)$. Hence in total we have
\begin{align*}
N^\times(2^h,m)&=
\begin{cases}
0,&\!\!\!\!h\le 2\\
(2^{h-1}-2)2^{m-1}+2^hN^\times(h,m-1),&\!\!\!\!h>2
\end{cases}\\
&=
\begin{cases}
0,&h\le 2\\
\frac{(2^{h-1}-2)(2^{hm}-2^m)}{2^h-2},&h>2,
\end{cases}\\
\intertext{and}
N^+(2^h,m)&=
\begin{cases}
2(2^{m-2}+N^+(1,m-1))&h=1\\
2^h(2^{m-1}+N^+(h,m-1))&h>1\\
\end{cases}\\
&=
\begin{cases}
(2m-1)2^{m-2}&h=1\\
\frac{2^{h(m-1)}+2^{h+m}-5\cdot2^{hm-1}}{2^h-2}&h>1.
\end{cases}
\end{align*}
\par
Table \ref{tab:complexity} lists the number of complex operations required to decode $\RM_q(1,m)$ for different $m$ and $q=2^h$. It also compares the complexity of our algorithm with the complexity of \cite[Algorithm 1]{Grant1998}. Notice that for $h=1$ only real additions are required, and its number is equal to half the indicated value.
It is apparent that Algorithm \ref{alg:Dec} has lower complexity than \cite[Algorithm 1]{Grant1998}. To be precise, our algorithm requires exactly one half of the multiplications required for \cite[Algorithm 1]{Grant1998}. For $h>1$, the ratio of the number of complex additions required for Algorithm \ref{alg:Dec} and \cite[Algorithm 1]{Grant1998} is approximately $(2^{h+1}+2^{h-1}-1)/2^{2h}$ and converges to this value as $m$ approaches infinity. We finally remark that the ML decoder for $\RM_4(1,m)$ in \cite{Jones1999} requires $(m+1)2^{2m+1}$ real additions. Hence it has higher complexity than Algorithm \ref{alg:Dec}. 

\begin{table}[t]
\centering
\caption{Number of complex operations required to decode $\RM_q(1,m)$}
\label{tab:complexity}
\begin{tabular}{c|c|rr|rr}\hline
\multicolumn{2}{c}{} & \multicolumn{2}{|c|}{Algorithm \ref{alg:Dec}} & \multicolumn{2}{c}{\cite[Algorithm 1]{Grant1998}}\\\hline
$m$ & $q$ & $N^+$ & $N^\times$ & $N^+$ & $N^\times$ \\\hline
4 & 2 & 28      & 0          & 32        & 0\\
  & 4 & 256     & 0          & 480       & 0\\
  & 8 & 1600    & 1360       & 5440      & 2720\\
  & 16& 11392   & 28080      & 74880     & 56160\\\hline
5 & 2 & 72      & 0          & 80       & 0\\
  & 4 & 1088    & 0          & 1984      & 0\\
  & 8 & 12928   & 10912      & 43648     & 21824\\
  & 16 & 182528  & 449376     & 1198336   & 898752\\\hline
6 & 2 & 176     & 0          & 192       & 0\\
  & 4 & 4480    & 0          & 8064      & 0\\
  & 8 & 103680  & 87360      & 349440    & 174720\\
  & 16& 2920960 & 7190208    & 19173888 & 14380416\\\hline
\end{tabular}
\end{table}

\section{Conclusion}

We have presented a simple maximum-likelihood decoder for the first-order generalized Reed--Muller code $\RM_q(1,m)$ that enables efficient decoding of certain OFDM codes with strictly bounded PMPR. It was shown that the presented algorithm has lower complexity than previously proposed maximum-likelihood decoding schemes for $\RM_q(1,m)$.

\bibliographystyle{ieeetr}
\bibliography{references}

\end{document}